\documentclass[10pt,a4paper]{article}

\usepackage[utf8]{inputenc}
\usepackage[english]{babel}

\begin{document}
\large

\newpage
\begin{center}
\LARGE{\bf On the Type of the Spin Polarization Dependence
of the Neutrino Mass and Charge}
\end{center}
\large
\vspace{0.1cm}
\begin{center}
{\bf Rasulkhozha S. Sharafiddinov}
\end{center}
\vspace{0.1cm}
\begin{center}
{\bf Institute of Nuclear Physics, Uzbekistan Academy of Sciences,
\\Tashkent, 100214 Ulugbek, Uzbekistan}
\end{center}
\vspace{0.1cm}

\begin{verse}
\noindent
{\bf Abstract.} Any non-zero component of charge implies the existence of 
a kind of inertial mass. Therefore, each of the existing types of the dipole
moments must arise as a consequence of the availability of a kind of charge.
From their point of view, the elastic scattering of completely longitudinally
(transversally) polarized neutrinos and antineutrinos by spinless nuclei is
discussed taking into account the Coulomb, weak and the united electroweak
masses and charges, and also the magnetic, anapole and the electric dipole
moments of incoming fermions with the weak currents. Interconversions of
neutrinos of different components have been investigated, at which a
particle chiral invariance is violated at the expense of the flip of its
spin. This becomes possible owing to an intimate connection between these
phenomena and characters of the structure of Dirac mass. Analysis of the
studied process cross sections assumed that both masses and charges of
longitudinal and transversal neutrinos are strictly unidentical.
\end{verse}

\vspace{0.8cm}
\noindent
{\bf 1 Introduction}
\vspace{0.4cm}

\noindent
In studying the nature of elementary particles such characteristics as the
mass and charge play a large role. At the same time, it is well known that
according to the hypothesis of field mass based on the classical theory of 
an extensive electron [1], a particle all the mass is purely electric.
Such a structure, however, encounters many problems. One of them states that
the charge distribution in the electron is not steady.

My investigation of the interaction of leptons and their neutrinos with the field 
of emission shows clearly [2,3] that if the neutrino corresponds to the electron 
$(\nu=\nu_{e}),$ its full electric charge $e_{\nu}^{full}$ and magnetic moment 
$\mu_{\nu}^{full}$ appear [4] owing to the Coulomb rest mass $m_{\nu}^{E}$ 
and behave as 
\begin{equation}
e_{\nu}^{full}=
-\frac{3eG_{F}m_{\nu}^{2}}{4\pi^{2}\sqrt{2}}, \, \, \, \, e=|e|,
\label{1}
\end{equation}
\begin{equation}
\mu_{\nu}^{full}=
\frac{3eG_{F}m_{\nu}}{8\pi^{2}\sqrt{2}}, \, \, \, \, m_{\nu}=m_{\nu}^{E}.
\label{2}
\end{equation}

This picture reflects the fact that each of the available types of charges
says in favor of the availability of a kind of inertial mass. Thereby, the
possibility of the existence of the united rest mass $m_{\nu}^{U}$ and charge
$e_{\nu}^{U}$ for the neutrino equal to its all the mass and charge is not 
excluded. One can define their structure in the form [5]
\begin{equation}
m_{\nu}^{U}=m_{\nu}^{E}+m_{\nu}^{W}+m_{\nu}^{S}+...,
\label{3}
\end{equation}
\begin{equation}
e_{\nu}^{U}=e_{\nu}^{E}+e_{\nu}^{W}+e_{\nu}^{S}+....
\label{4}
\end{equation}
Here the indices $E,$ $W$ and $S$ correspond to the electric, weak and the
strong components of the neutrino mass and charge. They constitute herewith
the harmony of forces of a different nature. Therefore, the charge
distribution of the light lepton must be steady [6].

Such a steadiness of matter says about the compound structure [7] of charge
quantization law and thereby testifies in favor of that any non-zero 
component of the electric charge implies the existence of a kind of dipole 
moment. In other words, Dirac $F_{1\nu}(q^{2})$ and Pauli $F_{2\nu}(q^{2})$ 
form factors contain [6] the statical as well as the dynamical components:
\begin{equation}
F_{i\nu}(q^{2})=f_{i\nu}(0)+A_{i\nu}(\vec{q^{2}})+...,
\label{5}
\end{equation}
where $f_{i\nu}(0)$ give the normal sizes of charge and moment,
$A_{i\nu}(\vec{q^{2}})$ characterize the momentum dependence of leptonic
current vector parts. The terms $f_{i\nu}(0)$ and $A_{i\nu}(\vec{q^{2}})$
are responsible for the first and second Born approximations. In these
circumstances, the form factors $F_{1\nu}(q^{2})$ and $F_{2\nu}(q^{2})$
at $q^{2}=0$ define the full static values of the neutrino electric
charge and moment:
\begin{equation}
F_{1\nu}(0)=e_{\nu}^{full}=e_{\nu}^{norm}+e_{\nu}^{anom}+...,
\label{6}
\end{equation}
\begin{equation}
F_{2\nu}(0)=\mu_{\nu}^{full}=\mu_{\nu}^{norm}+\mu_{\nu}^{anom}+....
\label{7}
\end{equation}

Of course, the electric mass and charge of a Dirac neutrino correspond to
the most diverse forms of the same regularity of its Coulomb nature. By this
reason we conclude [4] that the neutrino electric mass $m_{\nu}^{E}$ includes
as well as the normal $m_{\nu}^{norm}$ and 
anomalous $m_{\nu}^{anom}$ components:
\begin{equation}
m_{\nu}^{E}=m_{\nu}^{norm}+m_{\nu}^{anom}+....
\label{8}
\end{equation}

However, it is known [8,9] that the neutrino interaction with virtual 
photons is described by the vertex operator
$$\Gamma_{\mu}(p,p')=\overline{u}(p',s')[\gamma_{\mu}F_{1\nu}(q^{2})-
i\sigma_{\mu\lambda}q_{\lambda}F_{2\nu}(q^{2})+$$
\begin{equation}
+\gamma_{5}\gamma_{\mu}G_{1\nu}(q^{2})-
i\gamma_{5}\sigma_{\mu\lambda}q_{\lambda}G_{2\nu}(q^{2})]u(p,s).
\label{9}
\end{equation}
Here $\sigma_{\mu\lambda}=[\gamma_{\mu},\gamma_{\lambda}]/2,$ $q=p-p',$
$p(s)$ and $p'(s')$ denote the four-momentum (helicities) of the neutrino
before and after the emission, $G_{1\nu}(q^{2})$ and $G_{2\nu}(q^{2})$ are 
the current axial-vector parts.

Analysis of electroweak processes on nuclei assumed [10] that $F_{2\nu}$ and 
$G_{2\nu}$ must have the same size. Therefore, without loss of generality, 
we may write
\begin{equation}
G_{i\nu}(q^{2})=g_{i\nu}(0)+\Phi_{i\nu}(\vec{q^{2}})+...,
\label{10}
\end{equation}
where $g_{1\nu}(0)$ and $g_{2\nu}(0)$ are the normal components of 
the particle anapole [11] and electric dipole moments. The functions 
$\Phi_{i\nu}(\vec{q^{2}})$ characterize the anomalous behavior of
axial-vector form factors.

According to the correspondence principle, each term in (\ref{10}) as well 
as in (\ref{5}), corresponds to the definite approximation. This sight on 
the interaction axial-vector nature quality explains the fact that 
$G_{1\nu}(0)$ and $G_{2\nu}(0)$ give the full static sizes of the Dirac
particle anapole and electric dipole moments:
\begin{equation}
G_{1\nu}(0)=a_{\nu}^{full}=a_{\nu}^{norm}+a_{\nu}^{anom}+...,
\label{11}
\end{equation}
\begin{equation}
G_{2\nu}(0)=d_{\nu}^{full}=d_{\nu}^{norm}+d_{\nu}^{anom}+....
\label{12}
\end{equation}

Of them ${\it a_{\nu}^{full}}$ also can be measured experimentally [12].
For $d_{\nu}^{full}$ as well as for $e_{\nu}^{full}$ and $\mu_{\nu}^{full}$
there exist laboratory and cosmological restrictions [13,14].

The purpose of the present work is to discuss the problem of the neutrino
latent mass investigating its interaction with the field of electroweak emission
in the polarization type dependence. First of all we consider the elastic
scattering of longitudinal polarized massive Dirac neutrinos by nuclei
of the electric $(Z)$ and weak $(Z_{W})$ charges
\begin{equation}
\nu_{L,R}({\bar \nu}_{R,L})+A(Z, Z_{W})\stackrel{\gamma,Z}
{\rightarrow}\nu'({\bar \nu'})+ A(Z, Z_{W})
\label{13}
\end{equation}
at the account of their Coulomb, weak and the united electroweak masses
and currents. Next, all they will be reanalyzed for the transversal case
of an incoming particle polarization. In conclusion we present some
implications implied from these considerations.

\vspace{0.8cm}
\noindent
{\bf 2 Interaction of longitudinal polarized neutrinos with the field
of a nucleus}
\vspace{0.4cm}

\noindent
From our earlier developments, we find that the matrix elements of elastic
scattering of arbitrary polarized neutrinos on the nucleus electric and weak 
charges in the first Born approximation must have the following structure:
$$M^{E}_{fi}=\frac{4\pi\alpha}{q_{E}^{2}}\overline{u}(p_{E}',s')
[\gamma_{\mu}f_{1\nu}^{E}(0)-
i\sigma_{\mu\lambda}q_{\lambda E}f_{2\nu}^{E}(0)+$$
\begin{equation}
+\gamma_{5}\gamma_{\mu}g_{1\nu}^{E}(0)-
i\gamma_{5}\sigma_{\mu\lambda}q_{\lambda E}g_{2\nu}^{E}(0)]u(p_{E},s)
J_{\mu}^{\gamma}(q_{E}),
\label{14}
\end{equation}
\begin{equation}
M^{W}_{fi}=
\frac{G_{F}}{\sqrt{2}}\overline{u}(p_{W}',s')\gamma_{\mu}(g_{V_{\nu}}+
\gamma_{5}g_{A_{\nu}})u(p_{W},s)J_{\mu}^{Z}(q_{W}).
\label{15}
\end{equation}
Here $\nu=\nu_{L,R}=\nu_{eL,R},$ $q_{E}=p_{E}-p_{E}',$ $q_{W}=p_{W}-p_{W}',$
$p_{E}(p_{W})$ and $p_{E}'(p_{W}')$ imply the four-momentum of the neutrino
before and after the electric (weak) emission, $f_{i\nu}^{E}$ and $g_{i\nu}^{E}$ 
characterize the Coulomb $m_{\nu}^{E}$ mass dependence of the neutrino form 
factors, $J_{\mu}^{x}$ are the nuclear photon $(x=\gamma)$ and weak $(x=Z)$ 
currents [15], $g_{V_{\nu}}$ and $g_{A_{\nu}}$ denote the corresponding 
constants of the purely weak interaction vector and axial-vector components.

As seen from (\ref{14}) and (\ref{15}), in the case of exchange by the photon,
only the Coulomb mass is responsible for the electric scattering. Insofar as
$m_{\nu}^{W}$ is concerned, it leads to the corresponding weak interaction.

It appears that on the basis of the standard definition
\begin{equation}
\frac{d\sigma_{E,W}(s,s')}{d\Omega}=
\frac{1}{16\pi^{2}}|M^{E}_{fi}+M^{W}_{fi}|^{2}
\label{16}
\end{equation}
one can establish an explicit form of the studied process cross sections. 
It is not excluded, however, that any Dirac particle possesses simultaneously 
both electric and weak [16] masses. In other words, the neutrino interaction 
with the field of emission explained by its electroweak $m_{\nu}^{EW}$ mass 
arises at the expense of exchange simultaneously both by the photon and by 
the weak boson. From this point of view, the interference $(I)$ between the 
interactions (\ref{14}) and (\ref{15}) can be expressed as follows:
$$ReM^{E}_{fi}M^{*W}_{fi}=
\frac{4\pi\alpha G_{F}}{\sqrt{2}q_{EW}^{2}}
Re\Lambda_{EW}\Lambda_{EW}'[\gamma_{\mu}f_{1\nu}^{I}(0)-$$
$$-i\sigma_{\mu\lambda}q_{\lambda EW}f_{2\nu}^{I}(0)+
\gamma_{5}\gamma_{\mu}g_{1\nu}^{I}(0)-$$
\begin{equation}
-i\gamma_{5}\sigma_{\mu\lambda}q_{\lambda EW}g_{2\nu}^{I}(0)]
\gamma_{\mu}(g_{V_{\nu}}+\gamma_{5}g_{A_{\nu}})
J_{\mu}^{\gamma}(q_{EW})J_{\mu}^{Z}(q_{EW}),
\label{17}
\end{equation}
where the interaction of each of the currents $f_{i\nu}^{I}$ and $g_{i\nu}^{I}$
with the united field of emission of the photon and weak boson is explained
by the electroweak structures of mass and charge. They are of course strictly
interference. Here it is also necessary to keep in mind that
$$q_{EW}=p_{EW}-p_{EW}', \, \, \, \, m_{\nu}^{EW}=m_{\nu}^{E}+m_{\nu}^{W},$$
$$\Lambda_{EW}=u(p_{EW},s)\overline{u}(p_{EW},s), \, \, \, \, \Lambda_{EW}'=
u(p_{EW}',s')\overline{u}(p_{EW}',s').$$

According to these data, the cross section of the process (\ref{13}) at the 
account of longitudinal polarization of both incoming and outgoing fermions 
is written in the form
\begin{equation}
d\sigma_{E,W}^{V,A}(\theta_{E,W},s,s')=
d\sigma_{E}^{V,A}(\theta_{E},s,s')+
d\sigma_{I}^{V,A}(\theta_{EW},s,s')+
d\sigma_{W}^{V,A}(\theta_{W},s,s'),
\label{18}
\end{equation}
where the contribution of purely electric mass has the size
$$\frac{d\sigma_{E}^{V,A}(\theta_{E},s,s')}{d\Omega}=
\frac{1}{2}\sigma^{E}_{o}(1-\eta^{2}_{E})^{-1}\{(1+ss')[f_{1\nu}^{E}+$$
$$+2\lambda_{c}s\sqrt{1-\eta_{E}^{2}}g_{1\nu}^{E}]f_{1\nu}^{E}+$$
$$+\eta^{2}_{E}(1-ss')[(f_{1\nu}^{E})^{2}+
4(m_{\nu}^{E})^{2}(1-\eta^{-2}_{E})^{2}(f_{2\nu}^{E})^{2}]
tg^{2}\frac{\theta_{E}}{2}-$$
$$-8s(E_{\nu}^{E})^{2}(1-ss')(1-\eta_{E}^{2})^{3/2}f_{2\nu}^{E}g_{2\nu}^{E}
tg^{2}\frac{\theta_{E}}{2}+$$
$$+(1-\eta^{2}_{E})[(1+ss')(g_{1\nu}^{E})^{2}+$$
\begin{equation}
+4(E_{\nu}^{E})^{2}(1-ss')(g_{2\nu}^{E})^{2}
tg^{2}\frac{\theta_{E}}{2}]\}
F_{E}^{2}(q_{E}^{2}).
\label{19}
\end{equation}

As a consequence of the availability of the united electroweak mass in particles, the 
second term characterizes the interference of electric and weak interactions: 
$$\frac{d\sigma_{I}^{V,A}(\theta_{EW},s,s')}{d\Omega}=
\frac{1}{2}\rho_{EW}\sigma^{EW}_{o}(1-\eta^{2}_{EW})^{-1}g_{V_{\nu}}
\{(1+ss')[1-$$
$$-\lambda_{c}s\frac{g_{A_{\nu}}}{g_{V_{\nu}}}\sqrt{1-
\eta^{2}_{EW}}][f_{1\nu}^{I}+
\lambda_{c}s\sqrt{1-\eta_{EW}^{2}}g_{1\nu}^{I}]+$$
\begin{equation}
+\eta^{2}_{EW}(1-ss')f_{1\nu}^{I}
tg^{2}\frac{\theta_{EW}}{2}\}F_{I}(q_{EW}^{2}).
\label{20}
\end{equation}

The corresponding cross section for the process going at the expense
of the neutrino purely weak rest mass behaves as
$$\frac{d\sigma_{W}^{V,A}(\theta_{W},s,s')}{d\Omega}=
\frac{G_{F}^{2}(E_{\nu}^{W})^{2}}{16\pi^{2}}
\{g_{V_{\nu}}^{2}[(1+ss')
cos^{2}\frac{\theta_{W}}{2}+$$
$$+\eta_{W}^{2}(1-ss')
sin^{2}\frac{\theta_{W}}{2}]+
g_{A_{\nu}}^{2}(1+ss')(1-\eta_{W}^{2})
cos^{2}\frac{\theta_{W}}{2}-$$
\begin{equation}
-2\lambda_{c}sg_{V_{\nu}}g_{A_{\nu}}(1+ss')
\sqrt{1-\eta_{W}^{2}}
cos^{2}\frac{\theta_{W}}{2}\}F_{W}^{2}(q_{W}^{2}).
\label{21}
\end{equation}
Here we must have in view of that
$$\sigma_{o}^{E}=
\frac{\alpha^{2}cos^{2}(\theta_{E}/2)}
{4(E_{\nu}^{E})^{2}(1-\eta^{2}_{E})sin^{4}(\theta_{E}/2)}, \, \, \, \,
\rho_{EW}=\frac{G_{F}q_{EW}^{2}}{2\pi\sqrt{2}\alpha},$$
$$\sigma_{o}^{EW}=
\frac{\alpha^{2}cos^{2}(\theta_{EW}/2)}
{4(E_{\nu}^{EW})^{2}(1-\eta^{2}_{EW})sin^{4}(\theta_{EW}/2)}, \, \, \, \,
\eta_{E}=\frac{m_{\nu}^{E}}{E_{\nu}^{E}},$$
$$\eta_{EW}=\frac{m_{\nu}^{EW}}{E_{\nu}^{EW}}, \, \, \, \,
\eta_{W}=\frac{m_{\nu}^{W}}{E_{\nu}^{W}}, \, \, \, \,
F_{E}(q_{E}^{2})=ZF_{c}(q_{E}^{2}),$$
$$F_{I}(q_{EW}^{2})=ZZ_{W}F_{c}^{2}(q_{EW}^{2}), \, \, \, \,
F_{W}(q_{W}^{2})=Z_{W}F_{c}(q_{W}^{2}),$$
$$Z_{W}=\frac{1}{2}\{\beta_{V}^{(0)}(Z+N)+\beta_{V}^{(1)}(Z-N)\},$$
$$A=Z+N, \, \, \, \, M_{T}=\frac{1}{2}(Z-N),$$
where $\theta_{E},$ $\theta_{EW}$ and $\theta_{W}$ are the polar angles
in the electric, electroweak and purely weak scattering at the neutrino
energies $E_{\nu}^{E},$ $E_{\nu}^{EW}$ and $E_{\nu}^{W},$ the functions
$F_{c}(q_{E}^{2}),$ $F_{c}(q_{EW}^{2})$ and $F_{c}(q_{W}^{2})$ describe
the charge $(F_{c}(0)=1)$ form factors of a nucleus in these three processes,
$M_{T}$ is the projection of its isospin $T,$ constants $\beta_{V}^{(0)}$
and $\beta_{V}^{(1)}$ correspond to the isoscalar and isovector components
of the nuclear vector weak current.

The indices $V$ and $A$ imply the presence simultaneously of both vector and 
axial-vector parts of leptonic photon and weak currents. Therefore, any of the 
expressions (\ref{19})-(\ref{21}) for the neutrino $(\lambda_{c}=+1)$ and the 
antineutrino $(\lambda_{c}=-1)$ is different which becomes possible owing to 
the interference of the interaction vector and axial-vector components.

The terms $(1+ss')$ and $(1-ss')$ characterize the scattering with conservation 
$(s'= s)$ and change $(s'=-s)$ of helicities of incoming left $(s=-1)$- and 
right $(s=+1)$-polarized particles. Under such circumstances, the cross section 
(\ref{18}) is convenient to replace by the summed size
\begin{equation}
d\sigma_{E,W}^{V,A}(\theta_{E,W},s)=
d\sigma_{E,W}^{V,A}(\theta_{E,W},s'=s)+
d\sigma_{E,W}^{V,A}(\theta_{E,W},s'=-s).
\label{22}
\end{equation}

The compound structures of both terms of (\ref{22}) testify in favor of that
the neutrino charge $f_{1\nu}$ leads to the scattering either with or without 
flip of its spin. The anapole $g_{1\nu}$ does not change a particle helicity. 
In contrast to this, the vector $f_{2\nu}$ and axial-vector $g_{2\nu}$ moments 
are responsible only for the flip of the neutrino spin. Of course, our formulas 
can also confirm the fact that the helicity of the neutrino of large energy 
$(E_{\nu}\gg m_{\nu})$ is not changed.

However, as known, the right-handed neutrino encounters the problem which
states that a chiral symmetry characterized a massive particle does not exist.
Therefore, it appears that the neutrinos have no neither the electric, weak nor
any other mass. But this is not quite so. The point is that at the helicity
conservation, a particle chirality is not changed even if it possesses a
non-zero rest mass. In other words, the longitudinal neutrino chirality
can be violated at the expense of mass, charge, magnetic and electric 
dipole moments, because they lead to the flip of its spin.

In the absence of one of the currents, $V$ or $A,$ any of (\ref{19})-(\ref{21})
not only for the particle and the antiparticle, but also for the left- and
right-handed neutrinos coincides. Such an equality takes place as well
as in the low energy limits of the corresponding processes.

\vspace{0.8cm}
\noindent
{\bf 3 Transversal polarized neutrino scattering by nuclei
of electroweak charges}
\vspace{0.4cm}

\noindent
Owing to an intimate connection between the mass of a particle and its physical 
nature, any massive neutrino has the longitudinal as well as the transversal 
polarization. Here an important circumstance is the fact that the same neutrino 
must not be simultaneously both a longitudinal and a transversal fermion. There 
exists, however, the possibility that the longitudinal polarized neutrinos in 
the elastic scattering on a nucleus can be converted into the transversal 
polarized ones and vice versa [17].

Returning to (\ref{14})-(\ref{17}), we establish the cross section of the process 
(\ref{13}) for the transversal case of the neutrino polarization which one can 
present as follows:
\begin{equation}
d\sigma_{E,W}^{V,A}(\theta_{E,W},\varphi,s,s')=
d\sigma_{E}^{V,A}(\theta_{E},\varphi,s,s')+
d\sigma_{I}^{V,A}(\theta_{EW},\varphi,s,s')+
d\sigma_{W}^{V,A}(\theta_{W},\varphi,s,s').
\label{23}
\end{equation}
Here to the contribution of purely electric mass responds the expression
$$\frac{d\sigma_{E}^{V,A}(\theta_{E},\varphi,s,s')}{d\Omega}=
\frac{1}{2}\sigma^{E}_{o}(1-\eta^{2}_{E})^{-1}
\{((1+ss')\alpha_{T}cos^{2}\frac{\varphi}{2}+$$
$$+(1-ss')\alpha^{*}_{T}sin^{2}\frac{\varphi}{2})(f_{1\nu}^{E})^{2}+
\eta^{2}_{E}((1+ss')\gamma_{T}sin^{2}\frac{\varphi}{2}-$$
$$-(1-ss')\gamma^{*}_{T}cos^{2}\frac{\varphi}{2})
[(f_{1\nu}^{E})^{2}+4(m_{\nu}^{E})^{2}(1-\eta^{-2}_{E})^{2}(f_{2\nu}^{E})^{2}]
tg^{2}\frac{\theta_{E}}{2}+$$
$$+2\lambda_{c}s\eta_{E}\sqrt{1-\eta_{E}^{2}}
((1+ss')sin^{2}\frac{\varphi}{2}-$$
$$-(1-ss')cos^{2}\frac{\varphi}{2})
\gamma^{*}_{T}f_{1\nu}^{E}g_{1\nu}^{E}
tg\frac{\theta_{E}}{2}+$$
$$+(1-\eta^{2}_{E})((1+ss')\alpha^{*}_{T}sin^{2}\frac{\varphi}{2}+
(1-ss')\alpha_{T}cos^{2}\frac{\varphi}{2})(g_{1\nu}^{E})^{2}+$$
$$+4(E_{\nu}^{E})^{2}(1-\eta^{2}_{E})
((1+ss')\gamma^{*}_{T}cos^{2}\frac{\varphi}{2}-$$
\begin{equation}
-(1-ss')\gamma_{T}sin^{2}\frac{\varphi}{2})(g_{2\nu}^{E})^{2}
tg^{2}\frac{\theta_{E}}{2}\}F_{E}^{2}(q_{E}^{2}).
\label{24}
\end{equation}

The interference process cross section originated at the expense of the united 
electroweak mass of transversal polarized neutrinos has the following structure:
$$\frac{d\sigma_{I}^{V,A}(\theta_{EW},\varphi,s,s')}{d\Omega}=
\frac{1}{2}\rho_{EW}\sigma^{EW}_{o}(1-\eta^{2}_{EW})^{-1}g_{V_{\nu}}
\{((1+ss')\alpha_{T}cos^{2}\frac{\varphi}{2}+$$
$$+(1-ss')\alpha^{*}_{T}sin^{2}\frac{\varphi}{2})f_{1\nu}^{I}+$$
$$+\eta_{EW}^{2}[((1+ss')\gamma_{T}sin^{2}\frac{\varphi}{2}-
(1-ss')\gamma^{*}_{T}cos^{2}\frac{\varphi}{2})tg\frac{\theta_{EW}}{2}+$$
$$+\lambda_{c}s\frac{g_{A_{\nu}}}{g_{V_{\nu}}}\eta_{EW}^{-1}
\sqrt{1-\eta^{2}_{EW}}((1+ss')sin^{2}\frac{\varphi}{2}-$$
$$-(1-ss')cos^{2}\frac{\varphi}{2})\gamma^{*}_{T}]
f_{1\nu}^{I}tg\frac{\theta_{EW}}{2}-$$
$$-\lambda_{c}s\eta_{EW}\sqrt{1-\eta^{2}_{EW}}
[((1+ss')sin^{2}\frac{\varphi}{2}-$$
$$-(1-ss')cos^{2}\frac{\varphi}{2})
\gamma^{*}_{T}tg\frac{\theta_{EW}}{2}+$$
$$+\lambda_{c}s\frac{g_{A_{\nu}}}{g_{V_{\nu}}}\eta_{EW}^{-1}
\sqrt{1-\eta^{2}_{EW}}((1+ss')\alpha^{*}_{T}sin^{2}\frac{\varphi}{2}+$$
\begin{equation}
+(1-ss')\alpha_{T}cos^{2}\frac{\varphi}{2})]g_{1\nu}^{I}\}F_{I}(q_{EW}^{2}).
\label{25}
\end{equation}

The contribution explained by the transversal polarized neutrino purely
weak rest mass is written in the form
$$\frac{d\sigma_{W}^{V,A}(\theta_{W},\varphi,s,s')}{d\Omega}=
\frac{G_{F}^{2}(E_{\nu}^{W})^{2}}{16\pi^{2}}
\{g_{V_{\nu}}^{2}[((1+ss')\alpha_{T}cos^{2}\frac{\varphi}{2}+$$
$$+(1-ss')\alpha^{*}_{T}sin^{2}\frac{\varphi}{2})
ctg^{2}\frac{\theta_{W}}{2}+$$
$$+\eta_{W}^{2}((1+ss')\gamma_{T}sin^{2}\frac{\varphi}{2}-
(1-ss')\gamma^{*}_{T}cos^{2}\frac{\varphi}{2})]
sin^{2}\frac{\theta_{W}}{2}+$$
$$+g_{A_{\nu}}^{2}(1-\eta_{W}^{2})
((1+ss')\alpha^{*}_{T}sin^{2}\frac{\varphi}{2}+$$
$$+(1-ss')\alpha_{T}cos^{2}\frac{\varphi}{2})
cos^{2}\frac{\theta_{W}}{2}-$$
$$-2\lambda_{c}sg_{V_{\nu}}g_{A_{\nu}}\eta_{W}
\sqrt{1-\eta_{W}^{2}}
((1+ss')sin^{2}\frac{\varphi}{2}-$$
\begin{equation}
-(1-ss')cos^{2}\frac{\varphi}{2})\gamma^{*}_{T}
sin\frac{\theta_{W}}{2}cos\frac{\theta_{W}}{2}\}
F_{W}^{2}(q_{W}^{2}),
\label{26}
\end{equation}
where it has been accepted that
$$ \alpha_{T}=1-2(1-4sin^{2}\frac{\varphi}{2})
sin^{2}\frac{\varphi}{2},$$
$$ \alpha^{*}_{T}=
1+2(1-4sin^{2}\frac{\varphi}{2})cos^{2}\frac{\varphi}{2},$$
$$ \gamma_{T}=1+2cos^{2}\frac{\varphi}{2},\,\,\,\, \gamma^{*}_{T}=
1-2cos^{2}\frac{\varphi}{2}.$$
Here $\varphi$ is the azimuthal angle.

As well as in (\ref{18}), each term in (\ref{23}) contains the contributions
of vector and axial-vector interactions, and also the contributions of their
interference between themselves owing to which, the neutrino and antineutrino
scattering cross sections are different.

Furthermore, if taken into account the availability of the multipliers
$(1+ss')$ and $(1-ss')$ in (\ref{24})-(\ref{26}), we can present
(\ref{23}) in the form
\begin{equation}
d\sigma_{E,W}^{V,A}(\theta_{E,W},\varphi,s)=
d\sigma_{E,W}^{V,A}(\theta_{E,W},\varphi,s'=s)+
d\sigma_{E,W}^{V,A}(\theta_{E,W},\varphi,s'=-s).
\label{27}
\end{equation}

An explicit expressions for both terms of (\ref{27}) have the most
diverse structures. From their point of view, in the case of the neutrino
transversal polarization, the processes with or without change of incoming
particle helicities must go as a consequence not only of charge, but also
of any dipole moment. It is of course not excluded that the flip of the
transversal neutrino spin which arises at the expense of mass, charge,
magnetic, anapole and electric dipole moments can explain the possible 
violation of its chirality.

The absence of one of the currents, $V$ or $A,$ implies that each of
(\ref{24})-(\ref{26}) for the neutrino and the antineutrino as well as for
the left- and right-handed particles is not different. Such a coincidence
takes place even in the low energy limits of the corresponding types
of interactions.

\vspace{0.8cm}
\noindent
{\bf 4 Conclusion}
\vspace{0.4cm}

\noindent
In conformity with the laws of neutrino nature, the presence of any type 
of charge implies the existence of a kind of inertial mass [5]. Such a 
duality of matter says about the steadiness of charge distribution in the 
neutrino and thereby testifies in favor of that each of all possible types 
of the dipole moments arises as a consequence of the availability of a kind 
of charge [6]. Therefore, to reanalyze these features and discuss their some 
implications, we have established the compound structures of the differential 
cross sections describing the elastic scattering of completely longitudinally 
(transversally) polarized neutrinos and antineutrinos by spinless nuclei taking 
into account the Coulomb, weak and the united electroweak masses and charges, 
and also the magnetic, anapole and the electric dipole moments of incoming 
fermions with the weak currents.

They state that if neutrinos are of longitudinal polarized, their charge
answers to the elastic scattering either with or without flip of the spin.
The anapole is responsible only for a particle helicity conservation.
Unlike this, both magnetic and electric dipole moments lead to the
interconversion of neutrinos of different components. However, in the
transversal case of the neutrino polarization, each of these processes 
can originate through the interaction with the field of emission of 
charge as well as of any dipole moment.

The existence of interconversions $\nu_{L}\leftrightarrow \nu_{R}$ and
${\bar \nu}_{R}\leftrightarrow {\bar \nu_{L}}$ is incompatible with
chiral invariance. These transitions, however, take place owing to the rest
mass dependence of the behavior of neutrinos. At our sight, this connection
implies that a particle chirality is violated at the expense of the flip
of its helicity.

In the case of both longitudinal and transversal polarization of the neutrino,
the process (\ref{13}) is described by the three differential cross sections
corresponding to the electric, weak and the united electroweak masses and
charges. These cross sections can be defined simultaneously for the same
energy if all the three momentum transfer have the space-like size.

One of the beautiful new features of our formulas is the indication to the
existence of different low energy limits for the same particle in the
interaction type dependence. They of course in the slow neutrino
scattering by nuclei behave as
\begin{equation}
E_{\nu}^{E}\rightarrow m_{\nu}^{E}, \, \, \, \,
E_{\nu}^{EW}\rightarrow m_{\nu}^{EW}, \, \, \, \,
E_{\nu}^{W}\rightarrow m_{\nu}^{W}.
\label{28}
\end{equation}

At these values, the cross sections (\ref{22}) and (\ref{27}) describing
the processes with longitudinal and transversal fermions are not different:
\begin{equation}
d\sigma_{E,W}^{V,A}(\theta_{E,W},s)=
d\sigma_{E,W}^{V,A}(\theta_{E,W},\varphi,s).
\label{29}
\end{equation}

For an arbitrary energy such a situation takes place when either vector
or axial-vector interactions are present:
\begin{equation}
d\sigma_{E,W}^{V}(\theta_{E,W},s)=
d\sigma_{E,W}^{V}(\theta_{E,W},\varphi,s),
\label{30}
\end{equation}
\begin{equation}
d\sigma_{E,W}^{A}(\theta_{E,W},s)=
d\sigma_{E,W}^{A}(\theta_{E,W},\varphi,s).
\label{31}
\end{equation}

But all the three equalities (\ref{29})-(\ref{31}) there exist only 
at the condition that a particle rest mass does not depend on the 
type of polarization.

Thus, if it turns out that at the availability of a non-zero mass, the
longitudinal polarized neutrino must be converted into the transversal
polarized one and vice versa [17], this will indicate to the existence 
of fundamental differences in the masses as well as in the charges of
longitudinal and transversal neutrinos.

Comparing (\ref{19}) with (\ref{24}), it is easy to observe the contribution
$sf_{1\nu}^{E}g_{1\nu}^{E}$ which is absent at the elastic scattering of 
longitudinal neutrinos on nuclei, but arises as a result of their transversal 
polarization. The term $sf_{2\nu}^{E}g_{2\nu}^{E}$ available in the longitudinal 
case of the neutrino polarization, does not appear at the transversal particle 
interaction. We can, therefore, conclude that the invariance of vector and 
axial-vector types of electroweak currents of longitudinal and transversal 
neutrinos concerning C, P and T, and also their combinations CP and CPT 
are different.

Finally, insofar as the spin polarization type dependence of the behavior
of massive Majorana neutrinos is concerned, this question together with
some aspects of the geometrical nature of inertial mass will be treated
in one of our further articles.

\vspace{0.8cm}
\noindent
{\bf References}
\begin{enumerate}
\item
E. Fermi, Rend. Lincei {\bf 31}, 184, 306 (1922); Phys. Zeit. {\bf 23}, 
340 (1922).
\item
R.S. Sharafiddinov, Dokl. Akad. Nauk Ruz. Ser. Math. Tehn. Estest. 
{\bf 7}, 25 (1998).
\item
R.S. Sharafiddinov, in Proc. Ukrain-Russian Grav. Conf. on Gravitation, 
Cosmology and Relativistic Astrophysics, Kharkov, November 8-11, 2000 
(Kharkov, Ukraine, 2000), p. 25.
\item
R.S. Sharafiddinov, Spacetime Subst. {\bf 1}, 176 (2000); hep-ph/0305009.
\item
R.S. Sharafiddinov, Spacetime Subst. {\bf 3}, 47 (2002); physics/0305008.
\item
R.S. Sharafiddinov, Spacetime Subst. {\bf 3}, 86 (2002); physics/0305009.
\item
R.S. Sharafiddinov, Spacetime Subst. {\bf 3}, 132 (2002); physics/0305014.
\item
M.A.B. Beg, W.J. Marciano and M. Ruderman, Phys. Rev. 
{\bf D 17}, 1395 (1978).
\item
W. Bernreuther and M. Suzuki, Rev. Mod. Phys.
{\bf 63}, 313 (1991).
\item
R.B. Begzhanov and R.S. Sharafiddinov, Mod. Phys. Lett. {\bf A 15}, (2000) 557; 
Izv. Russ. Acad. Nauk Ser. Fiz. {\bf 64}, 2221 (2000).
\item
Ya.B. Zel'dovich, JETP {\bf 33}, 1531 (1957); Ya.B. Zel'dovich 
and A.M. Perelomov, JETP {\bf 39}, 1115 (1960).
\item
M.J. Musolf and B.R. Holstein, Phys. Rev. {\bf D 43}, 1956 (1991).
\item
S. Davidson, B. Campbell and K.D. Bailey, Phys. Rev. {\bf D 43}, 2314 (1991).
\item
J.A. Morgan and D.B. Farrant, Phys. Lett. {\bf B 128}, 431 (1983).
\item
T.W. Donnelly and R.D. Peccei, Phys. Rep. {\bf 50}, 3 (1979).
\item
S. Weinberg, Phys. Rev. Lett. {\bf 29}, 388 (1972).
\item
R.S. Sharafiddinov, in Proc. Int. Conf. on Nuclear Physics, St-Petersburg, 
June 14-17, 2000 (St-Petersburg, 2000), p. 121.
\end{enumerate}

\end{document}


\large

\newpage
\begin{center}
\LARGE{\bf Addendum: On the Type of the Spin Polarization Dependence
of the Neutrino Mass and Charge}
\end{center}
\large
\vspace{0.1cm}
\begin{center}
{\bf Rasulkhozha S. Sharafiddinov}
\end{center}
\vspace{0.1cm}
\begin{center}
{\bf Institute of Nuclear Physics, Uzbekistan Academy of Sciences,
\\Tashkent, 100214 Ulugbek, Uzbekistan}
\end{center}
\vspace{0.1cm}

\begin{verse}
\noindent
{\bf Abstract.} We discuss the characteristic features of the latent nature 
of the neutrino mass, according to which, all components of leptonic current can 
appear in the interaction type dependence. Such a regularity, however, requires 
the modification of the some denotations in the corresponding formulas of the 
paper {\it On the type of the spin polarization dependence of the neutrino 
mass and charge} [1]. We will also include in a given letter the full version 
of the original article with necessary replacements of the structural variables. 
They of course do not change our implications.
\end{verse}
\vspace{0.4cm}

\noindent
In our original paper [1] the question has been raised as to whether there
exists any spin polarization type dependence of the neutrino mass, charge and
behavior. This sight on the nature of the neutrino is based logically on the
analysis of the possible processes of elastic scattering of longitudinal and
transversal polarized neutrinos $(\nu=\nu_{e})$ by spinless nuclei in 
the presence of the Coulomb $(K=E),$ weak $(K=W)$ and the united electroweak
$(K=EW)$ masses $m_{\nu}^{K},$ and also the charge $f_{1\nu},$ magnetic
$f_{2\nu},$ anapole $g_{1\nu}$ and electric dipole $g_{2\nu}$ moments of
incoming fermions with neutral currents. At the definition of their cross
sections, we used the matrix elements of transitions in the form
$$M^{E}_{fi}=\frac{4\pi\alpha}{q_{E}^{2}}\overline{u}(p_{E}',s')
[\gamma_{\mu}f_{1\nu}(0)-
i\sigma_{\mu\lambda}q_{\lambda E}f_{2\nu}(0)+$$
\begin{equation}
+\gamma_{5}\gamma_{\mu}g_{1\nu}(0)-
i\gamma_{5}\sigma_{\mu\lambda}q_{\lambda E}g_{2\nu}(0)]u(p_{E},s)
J_{\mu}^{\gamma}(q_{E}),
\label{1}
\end{equation}
\begin{equation}
M^{W}_{fi}=
\frac{G_{F}}{\sqrt{2}}\overline{u}(p_{W}',s')\gamma_{\mu}(g_{V_{\nu}}+
\gamma_{5}g_{A_{\nu}})u(p_{W},s)J_{\mu}^{Z}(q_{W}),
\label{2}
\end{equation}
where $q_{E}=p_{E}-p_{E}',$ $q_{W}=p_{W}-p_{W}',$ $p_{E}(p_{W})$ and
$p_{E}'(p_{W}')$ denote the four-momentum of a particle before and after
the electric (weak) scattering, $s$ and $s'$ are the helicities of these
states, $J_{\mu}^{\gamma}$ and $J_{\mu}^{Z}$ are the nuclear currents
in the studied processes, $g_{V_{\nu}}$ and $g_{A_{\nu}}$ correspond to
the vector and axial-vector components of leptonic weak neutral currents.

According to our description, a neutrino electric mass leads to its Coulomb 
scattering. At the exchange by the $Z$-boson, only the weak component of mass 
is responsible for the scattering.

In the framework of mass-charge duality [2], each of the existing types of 
charges testifies in favor of the availability of a kind of inertial mass.
Any particle of a Dirac nature can therefore have simultaneously both
electric and weak [3] masses.

Thus, in [1] the neutrino scattering by nuclei of an electroweak charge going 
at the expense of the interference between the interactions (\ref{1}) and 
(\ref{2}) is analytically described as
$$ReM^{E}_{fi}M^{*W}_{fi}=
\frac{4\pi\alpha G_{F}}{\sqrt{2}q_{EW}^{2}}
Re\rho_{EW}\rho_{EW}'[\gamma_{\mu}f_{1\nu}(0)-$$
$$-i\sigma_{\mu\lambda}q_{\lambda EW}f_{2\nu}(0)+
\gamma_{5}\gamma_{\mu}g_{1\nu}(0)-$$
\begin{equation}
-i\gamma_{5}\sigma_{\mu\lambda}q_{\lambda EW}g_{2\nu}(0)]
\gamma_{\mu}(g_{V_{\nu}}+\gamma_{5}g_{A_{\nu}})
J_{\mu}^{\gamma}(q_{EW})J_{\mu}^{Z}(q_{EW})
\label{3}
\end{equation}
and it is assumed additionally that
$$q_{EW}=p_{EW}-p_{EW}', \, \, \, \, m_{\nu}^{EW}=m_{\nu}^{E}+m_{\nu}^{W},$$
$$\rho_{EW}=u(p_{EW},s)\overline{u}(p_{EW},s),\,\, \, \, \rho_{EW}'=
u(p_{EW}',s')\overline{u}(p_{EW}',s').$$

Any component of leptonic current there exists at the availability of a kind
of charge [4], as follows from considerations of symmetry. If so, each of the 
interactions (\ref{1}) or (\ref{3}) states that the terms $f_{i\nu}$ and 
$g_{i\nu}$ responsible for the interference scattering are not of those form 
factors which arise as a result of particle Coulomb mass. The account of their
differences leads (\ref{1}) to the following:
$$M^{E}_{fi}=\frac{4\pi\alpha}{q_{E}^{2}}\overline{u}(p_{E}',s')
[\gamma_{\mu}f_{1\nu}^{E}(0)-
i\sigma_{\mu\lambda}q_{\lambda E}f_{2\nu}^{E}(0)+$$
\begin{equation}
+\gamma_{5}\gamma_{\mu}g_{1\nu}^{E}(0)-
i\gamma_{5}\sigma_{\mu\lambda}q_{\lambda E}g_{2\nu}^{E}(0)]u(p_{E},s)
J_{\mu}^{\gamma}(q_{E})
\label{4}
\end{equation}
and that, consequently, the interactions (\ref{3}) must have the form
$$ReM^{E}_{fi}M^{*W}_{fi}=
\frac{4\pi\alpha G_{F}}{\sqrt{2}q_{EW}^{2}}
Re\Lambda_{EW}\Lambda_{EW}'[\gamma_{\mu}f_{1\nu}^{I}(0)-$$
$$-i\sigma_{\mu\lambda}q_{\lambda EW}f_{2\nu}^{I}(0)+
\gamma_{5}\gamma_{\mu}g_{1\nu}^{I}(0)-$$
\begin{equation}
-i\gamma_{5}\sigma_{\mu\lambda}q_{\lambda EW}g_{2\nu}^{I}(0)]
\gamma_{\mu}(g_{V_{\nu}}+\gamma_{5}g_{A_{\nu}})
J_{\mu}^{\gamma}(q_{EW})J_{\mu}^{Z}(q_{EW}).
\label{5}
\end{equation}
Here $\Lambda_{EW}=\rho_{EW}$ and $\Lambda_{EW}'=\rho_{EW}'.$ Unlike
$f_{i\nu}^{E}$ and $g_{i\nu}^{E},$ the interference $(I)$ terms $f_{i\nu}^{I}$
and $g_{i\nu}^{I}$ characterize the united electroweak mass dependence
of the neutrino form factors.

Inserting (\ref{2}), (\ref{4}) and (\ref{5}) in
\begin{equation}
\frac{d\sigma_{E,W}(s,s')}{d\Omega}=
\frac{1}{16\pi^{2}}|M^{E}_{fi}+M^{W}_{fi}|^{2},
\label{6}
\end{equation}
we find for the elastic scattering cross sections the explicit expressions
which coincide with the corresponding formulas from the paper [1] if we
make there the following replacements:
$$f_{i\nu}\rightarrow f_{i\nu}^{t}, \, \, \, \,
g_{i\nu}\rightarrow g_{i\nu}^{t},$$
$$F_{EW}(q_{EW}^{2})\rightarrow F_{I}(q_{EW}^{2}), \, \, \, \,
\theta\rightarrow \theta_{K},$$
$$\sigma^{\nu}_{o}\rightarrow \sigma^{j}_{o}, \, \, \, \,
\rho\rightarrow \rho_{EW},$$
$$t=E,I, \, \, \, \, j=E,EW, \, \, \, \, K=E,EW,W.$$

Of them $\theta_{K}$ denotes the scattering angle in the mass type dependence, 
$F_{I}$ is the nuclear interference form factor, $\sigma^{j}_{o}$ describes 
the Mott cross section at the neutrino interaction in the presence of one 
of the masses, $m_{\nu}^{E}$ or $m_{\nu}^{EW},$ and $\rho_{EW}$ is the
electroweak interference coefficient.

Thus, a general picture is obtained of elastic scattering in which appears
an important part of compound structure of mass and charge.

Therefore, we will include in this letter the full version of the paper [1] 
with the above-noted modifications. They of course do not change our 
implications based on the analysis of earlier findings.

\newpage
\noindent
{\bf References}
\begin{enumerate}
\item
R.S. Sharafiddinov, Spacetime Subst. {\bf 3}, 134 (2002);
physics/0305015.
\item
R.S. Sharafiddinov, Spacetime Subst. {\bf 3}, 47 (2002);
physics/0305008.
\item
S. Weinberg, Phys. Rev. Lett. {\bf 29}, 388 (1972).
\item
R.S. Sharafiddinov, Spacetime Subst. {\bf 3}, 86 (2002);
physics/0305009.
\end{enumerate}